\documentclass[structabstract]{aa}

\usepackage{psfig}
\usepackage[english]{babel}
\usepackage{url}
\usepackage{amssymb}  
\usepackage{amsmath}
\usepackage{subfigure}
	
\usepackage{natbib}
\usepackage{graphicx}
\usepackage{subfigure}

\begin{document}
   \title{MHD modeling of coronal loops: injection of high-speed chromospheric flows }

   \author{A. Petralia
          \inst{1}
          \and
          F. Reale\inst{1,2}
          \and
          S. Orlando\inst{2}
          \and
          J. A. Klimchuk\inst{3}
              }

   \institute{Dipartimento di Fisica \& Chimica, Universit\`a di Palermo,
              Piazza del Parlamento 1, 90134 Palermo, Italy
           \and
             INAF-Osservatorio Astronomico di Palermo, Piazza del Parlamento 1, 90134 Palermo, Italy
             \and
             NASA Goddard Space Flight Center, Greenbelt, MD 20771, USA }

     \titlerunning{MHD modeling of coronal loops}

   \date{Received ; accepted }

% \abstract{}{}{}{}{}
% 5 {} token are mandatory

  \abstract
  % context heading (optional)
  % {} leave it empty if necessary
   {Observations reveal a correspondence between chromospheric type II spicules and bright upwardly moving fronts in the corona observed in the EUV band. However, theoretical considerations suggest that these flows are unlikely to be the main source of heating in coronal magnetic loops.}
  % aims heading (mandatory)
   {We investigate the propagation of high-speed chromospheric flows into coronal magnetic flux tubes, and the possible production of emission in the EUV band.}
  % methods heading (mandatory)
   {We simulate the propagation of a dense $10^4$ K chromospheric jet upwards along a coronal loop, by means of a 2-D cylindrical MHD model, including gravity, radiative losses, thermal conduction and magnetic induction. The jet propagates in a complete atmosphere including the chromosphere and a tenuous cool ($\sim 0.8$ MK) corona, linked through a steep transition region. In our reference model, the jet's initial speed is 70 km/s, its initial density is $10^{11}$ cm$^{-3}$, and the ambient uniform magnetic field is 10 G. We explore also other values of jet speed and density in 1-D, and of magnetic field in 2-D, and the jet propagation in a hotter ($\sim 1.5$ MK) background loop.}
  % results heading (mandatory)
   {While the initial speed of the jet does not allow it to reach the loop apex, a hot shock front develops ahead of it and travels to the other extreme of the loop. The shock front compresses the coronal plasma and heats it to about $10^6$ K. As a result, a bright moving front becomes visible in the 171 \AA\ channel of the SDO/AIA mission. This result generally applies to all the other explored cases, except for the propagation in the hotter loop.}
  % conclusions heading (optional), leave it empty if necessary
   {For a cool, low-density initial coronal loop, the post-shock plasma ahead of upward chromospheric flows might explain at least part of the observed correspondence between type II spicules and EUV emission excess.}


   \maketitle

\section{Introduction}

Type II spicules are very dynamic chromospheric events, with lifetimes ranging from 10 to 150~s and velocities between $50$ and $150$ km/s \citep{2007PASJ...59S.655D}. Cold plasma flows upwards at $10^{4}$ K,   and it has been proposed that a fraction of the plasma is heated to coronal temperatures (\citealt{2011Sci...331...55D}, \citealt{2011A&A...532L...1M}, \citealt{2012SoPh..280..425V}). For this reason and for the fact that they are very frequent events, it has been proposed that both ordinary spicules and type II spicules can give an important contribution to sustaining the high coronal temperatures (\citealt{1978ApJ...224L.143B}, \citealt{1978SoPh...57...49P}, \citealt{1982ApJ...255..743A}, \citealt{2004A&A...424..279T}, \citealt{2009ApJ...701L...1D}).

\cite{2012JGRA..11712102K} has discussed three observational tests of the hypothesis that most coronal plasma originates in the hot tips of spicules. These tests involve blue wing to line core intensity ratios, lower transition region to corona emission measure ratios, and blue wing to line core density ratios. The hypothesis fails all three tests by a wide margin (\citealt{2012JGRA..11712102K}, \citealt{2013ApJ...779....1T}, \citealt{2014ApJ...781...58P}).
The implication is that type II spicules inject at most a small fraction of the hot plasma that is present in the corona. Nonetheless, the evidence of correspondence between type II spicules and coronal EUV emission is well supported. It is therefore very important to understand how these flows are able to produce the observed emission.

One possible scenario can be that these jets produce shocks that propagate inside the loop, and heat and compress the plasma already present there. This may produce a transient excess EUV emission that would be observed. The fact that chromospheric material injected into a coronal flux tube
produces a shock, which may heat the coronal plasma, has been
known for some time (e.g., \citealt{1982ApJ...261..375K}). Here we address more specifically the shock heated material in front of the chromospheric jet.

We explore this scenario with a detailed MHD model, and test it quantitatively. A dense flow is triggered at one end of a magnetic flux tube linking two chromospheres through a tenuous corona. We study the propagation of this flow and investigate whether it drives signatures in EUV observations. In Section~\ref{sec:model} the model is described, we report on the simulations and the results in Section~\ref{sec:simul}, and discuss them in Sec.~\ref{sec:discuss}.

\section{The Model}
\label{sec:model}

We model the propagation of a dense chromospheric jet into a closed coronal loop anchored to the solar chromosphere. We simplify the model geometry to that of a straight magnetic flux tube linking two independent chromospheres. We image it as a vertical flux tube in 2D cylindrical geometry, i.e. the magnetic field is in the vertical direction. However, we maintain the gravity of a semicircular closed loop, i.e. the gravity is reduced as we move farther from the chromosphere and is zero at the loop apex, that corresponds to half the distance between the boundaries, at the middle of the vertical axis.

As our simulation strategy, we consider a reference model and then a number of other case studies with deviations from the reference model by one or more parameters.
For all the modeling, the initial condition is a complete coronal loop atmosphere. Since we want to test the suggestion that the corona comes from type II spicules, our reference loop atmosphere consists of  an initially very tenuous corona (EUV dark) linked to a thin chromospheric layer  through the usual steep transition region. Since we also want to understand the hot emission from type II spicules, even if it does not explain the bulk of the corona, we have run a simulation that uses typical conditions of an EUV bright corona. The atmosphere is immersed in a uniform magnetic field that links the two chromospheres. In the reference case, the intensity of the magnetic field implies values of the plasma $\beta \leq 1$ (see below), i.e. the plasma is mostly confined by the field. For completeness, we also consider a case of lower magnetic field and plasma $\beta \geq 1$.

In this atmosphere, a jet is injected from the bottom boundary upwards at time t = 0. Initially, the upflow has, therefore, chromospheric density and temperature. We are interested to study the propagation of the shock front that develops ahead of the upflow along the loop, and desire to maximize the observable effects from warm plasma emitting in the EUV band. So, we do not want the cold chromospheric material to fill the loop completely. Therefore, the speed is chosen to be relatively high, so as to produce a shock front, but not so high as to let the flow reach (and overcome) the loop apex. The resulting velocities of both the cold and the hot plasma are fully consistent with observations.

Our model solves the magnetohydrodynamic (MHD) equations for an ideal compressible plasma, in the following conservative form:

	\begin{equation}
	\dfrac{\partial \rho}{\partial t} + \triangledown \cdot (\rho \textbf{v}) = 0
	\end{equation}
	\begin{equation}
	\dfrac{\partial \rho \textbf{v}}{\partial t} + \triangledown \cdot (\rho \textbf{v} \textbf{v} - \textbf{B}\textbf{B} +p_{t}\textbf{I}) = \rho  \textbf{g}
	\end{equation}
	\begin{equation}
	\dfrac{\partial E}{\partial t} + \triangledown \cdot ( (E+p_{t} ) \textbf{v} - \textbf{B}( \textbf{v} \cdot \textbf{B}) ) = \rho  \textbf{v}  \cdot \textbf{g} - n_{e}n_{H} \Lambda(T) + H - \triangledown \cdot \textbf{F}_{c}
	\end{equation}
	\begin{equation}
	\dfrac{\partial  \textbf{B}}{\partial t} + \triangledown  \cdot ( \textbf{v} \textbf{B} - \textbf{B} \textbf{v} ) = 0
	\end{equation}
	\begin{equation}
	\triangledown \cdot \textbf{B} = 0
	\end{equation}
where
	\begin{equation}
	\rho = \mu m_{H} n_{H}
	\end{equation}
	\begin{equation}
	p_{t} = p +  \frac{ \textbf{B} \cdot \textbf{B}}{2}
	\end{equation}
	\begin{equation}
	E =\rho \epsilon+ \rho \frac{\textbf{v} \cdot \textbf{v}}{2} + \frac{ \textbf{B} \cdot \textbf{B}}{2}
	\end{equation}
	\begin{equation}
	\textbf{F}_{c} = \frac{ F_{sat}}{F_{sat} + |\textbf{F}_{class}|} \textbf{F}_{class}
	\end{equation}
	\begin{equation}
	\textbf{F}_{class} = k_{\parallel} \textbf{b} (\textbf{b} \cdot \bigtriangledown T ) + k_{\perp} ( \bigtriangledown T -
\textbf{b} (\textbf{b} \cdot \bigtriangledown T ) )
	\end{equation}
	\begin{equation}
	|\textbf{F}_{class}| = \sqrt{(\textbf{b} \cdot \bigtriangledown T)^{2}(k_{\parallel}^{2}-k_{\perp}^{2}) + k_{\perp}^{2}\bigtriangledown T^{2}}
	\end{equation}
	\begin{equation}
	F_{sat} =5 \Phi \rho c_{s}^{3}
	\end{equation}
where 	$ \mu = 1.265$ is the mean atomic mass (assuming solar metal abundances; \citealt{1989GeCoA..53..197A}), $m_{H}$ is the mass of hydrogen atom, $n_{H}$ is the hydrogen number density, $p_{t}$ is the total pressure, i.e. the sum of the thermal pressure and the magnetic pressure ( the factor $ 1/ \sqrt{4 \pi}$ is absorbed in the definition of $\textbf{B}$ ), $E$ is the total energy density, i.e. the sum of the thermal energy density ($\rho \epsilon$), the kinetic energy density and the magnetic energy, $\textbf{v}$ is the plasma velocity, $\textbf{g}$ is the solar gravity, $\Lambda(T)$ is the radiative loss function for optical thin plasma, $ \textbf{F}_{c}$ is the conductive flux, $H$ is a heating function, whose only role is to keep the unperturbed atmosphere in energy equilibrium, $c_{s}$ is the sound speed for an isothermal plasma, $\Phi$ is a free parameter ($<1$, \citealt{1984ApJ...277..605G}) that determines the degree of saturation of the thermal conduction; we set $\Phi = 0.9$ that corresponds to quite an efficient conduction. The radiative losses are computed according to version 7 of the CHIANTI code \citep{2012ApJ...744...99L}, assuming a density of $10^9$ cm$^{-3}$ and ionization equilibrium according to \citet{2009A&A...497..287D}. We do not account for the radiative losses of the chromospheric plasma, so we set $\Lambda(T)= 0$, as well as $ H = 0 $, for $T < 10^4$ K.

We complete this set with the equation of state for an ideal gas:
	\begin{equation}
	p = (\gamma -1)\rho \epsilon
	\end{equation}
	
The calculations are performed using the PLUTO code (\citealt{2007ApJS..170..228M}, \citealt{2012ApJS..198....7M}), a modular, Godunov-type code for astrophysical plasmas. The code provides a multiphysics, algorithmic modular environment particularly oriented toward the treatment of astrophysical flows in the presence of discontinuities as in
the case treated here. The code was designed to make efficient use of massive parallel computers using the message-passing interface (MPI) library for interprocessor communications. The MHD equations are solved using the MHD module available in PLUTO, configured to compute intercell fluxes with the Harten-Lax-Van Leer approximate Riemann solver, while second order in time is achieved using a Runge-Kutta scheme. A Van Leer limiter for the primitive variables is used. The evolution of the magnetic field is carried out adopting the constrained transport approach \citep{1999JCoPh.149..270B} that maintains the solenoidal condition ($\triangledown \cdot B = 0$) at machine accuracy. PLUTO includes optically thin radiative losses in a fractional step formalism \citep{2007ApJS..170..228M}, which preserves the 2nd order time accuracy, as the advection and source steps are at least of the 2nd order accurate; the radiative losses $\Lambda(T)$ values are computed at the temperature of interest using a table lookup/interpolation method. The thermal conduction is treated separately from advection terms through operator splitting. In particular, we adopted the super-time-stepping technique \citep{CNM:CNM950} which has been proved to be very effective to speed up explicit time-stepping schemes for parabolic problems. This approach is crucial when high values of plasma temperature are reached (as during flares). In fact,  the explicit scheme would require a very small time step because of the rather restrictive stability condition, i.e., $\bigtriangleup t \leqslant (\bigtriangleup x)^{2} /\eta$, where $\eta$ is the maximum diffusion coefficient \citep[e.g.,][]{2005A&A...444..505O,2008ApJ...678..274O}.

\subsection{The loop setup}

\begin{figure*}[!t]
    \subfigure[]{\includegraphics[scale= 0.5]{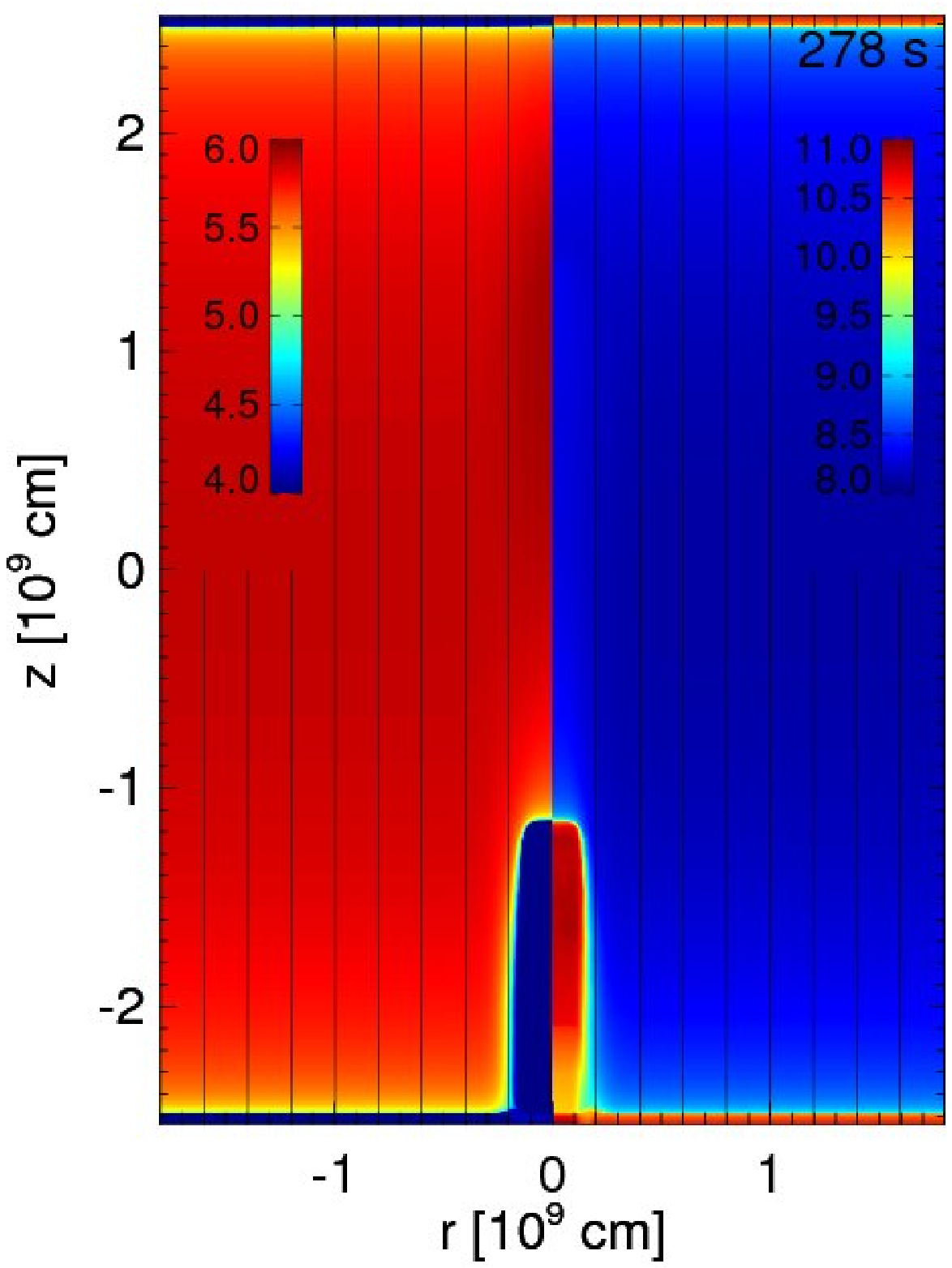}}
    \subfigure[]{\includegraphics[scale= 0.6]{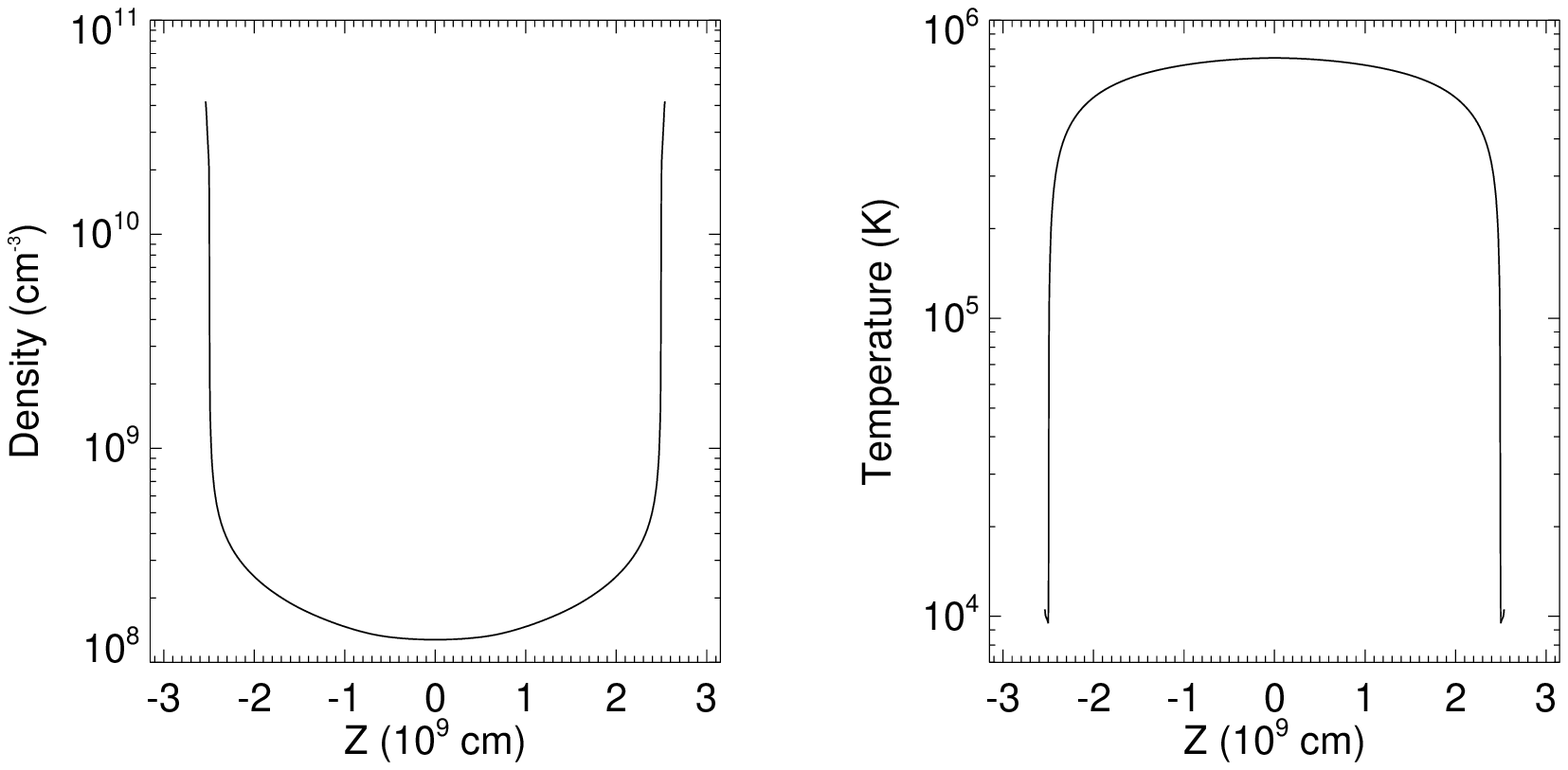}}
	\caption{Configuration of the computational domain: (a) maps of density (right, $cm^{-3}$) and temperature (left, $K$), in logarithmic scale, at time t = 278 s; (b) unperturbed density and temperature along the $z$ axis.}
	\label{fig:inicond}
	\end{figure*}

In the reference model, the computational domain is two-dimensional and cylindrical ($r, z$), and extends over $2L \sim 5.08 \times 10^{9}$ cm in the $z$ direction and $R_L \sim 1.8 \times 10^{9}$ cm in the $r$ direction, as shown in Fig.\ref{fig:inicond}a. As reference initial conditions, we consider a hydrostatic and relatively tenuous and cool unperturbed loop atmosphere,  (e.g. \citealt{2010LRSP....7....5R}); the density and temperature are set to be $n \sim 5 \times 10^{8}$ cm$^{-3}$ and  $T \sim 8 \times 10^{5}$ K, respectively, at the loop apex, i.e. at $z=0$, and $n \sim 10^{11}$ cm$^{-3}$ and  $T \sim 10^{4}$ K at base of the chromosphere, with the profiles shown in Fig.\ref{fig:inicond}b. The coronal heating is uniform with a value $H \sim 4.2 \times 10^{-5}$ erg cm$^{-3}$ s$^{-1}$. As alternative conditions, we consider also an initially hotter loop, where the density and temperature are $n \sim 10^{9}$ cm$^{-3}$ and  $T \sim 1.5 \times 10^{6}$ K, respectively, at the apex and a denser chromosphere  where  $n \sim 10^{12}$ cm$^{-3}$ and  $T \sim 10^{4}$ K at base of the chromosphere, with a uniform coronal heating of $H \sim 4.68 \times 10^{-4}$ erg cm$^{-3}$ s$^{-1}$.

The magnetic field is uniform in the $z$ direction and, in our reference configuration, has an intensity of 10 G. The field is therefore able to confine the plasma in the corona, as the plasma parameter $\beta$ is estimated:

\begin{equation}
\beta = \frac{p_{thermal}}{p_{magnetic}} = \frac{2n K_{B}T}{B^{2}/8\pi}\sim 0.1
\end{equation}
Alternatively, we also consider a magnetic field with an intensity of 3 G, that, together with a higher density of the jet, leads to a higher value of $\beta \geq 1$. We are aware that for a low value of $\beta$ the plasma evolution occurs mostly along the magnetic field lines, nevertheless for our reference model we equally consider a fully 2-D MHD description. This allows us to make a direct comparison with the case at higher $\beta$ which involves deviations from confinement. For the other simulations, the 2-D description is unnecessary and we consider a more efficient 1-D geometry.

Regarding the boundary conditions, we set axisymmetric conditions at $r = 0$, and reflective conditions at the $r = R_L$, where the magnetic field maintains the parallel component and reverses the perpendicular component (to the $z$ direction). At the $z$ boundaries we assume reflective conditions and the magnetic field maintains the perpendicular component and reverses the parallel component, except where the jet is defined. There we assume inflow boundary conditions.

The mesh of the 2-D domain is $z \times r = 1710 \times 160$ grid points and is uniformly-spaced all along the $z$-axis ($dz = 3.0 \times 10^6$ cm), and partially along the $r$-axis, from the center $r=0$ to a distance $r = R_U = 1.04 \times 10^9$ cm ($dr = 8.2 \times 10^6$ cm). For $r > R_U$ the spacing gradually expands to $dr = 5.2 \times 10^7$ cm; thus the boundary conditions are far enough to avoid any effect on the jet region. The uniformly high resolution along the $z$ direction allows us to describe accurately both the steep transition region and the moving shock front. To check for enough resolution in the steep gradient regions (e.g., \citealt{2013ApJ...770...12B}), we made a test at higher spatial resolution along $z$ ($dz = 0.5 \times 10^6$ cm) and ascertained that the results do not change except for a few details.

\section{The simulation}
\label{sec:simul}

\begin{figure}[!h]
	\centering
	\includegraphics[scale= 0.6]{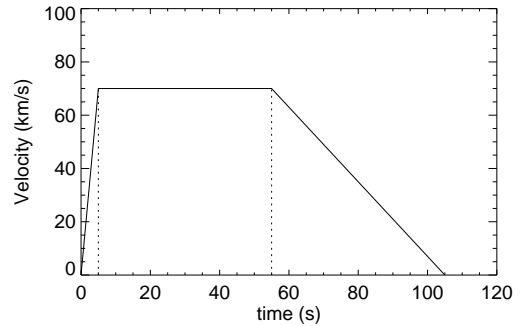}
	\caption{Temporal profile of the cold jet. }
	\label{fig:vel}
	\end{figure}

In the reference simulation, the jet is injected at the lower boundary of the domain with a velocity of 70 km/s, within a distance of $\delta r = 1.5 \times 10^{8}$ cm from the $z$ axis. We made some other tests with different jet velocities, i.e. 30, 50 and 90 km/s. Since spicules are observed to have a lifetime of a few minutes at most, for all simulations the flow is assumed to have a finite duration, described by the time profile in Fig.\ref{fig:vel}.
The bulk of the injection lasts 50 s. We set a linear rise and decrease of the flow speed that avoid numerical problems due to boundary conditions. The total duration of the flow is about 2 minutes. We do not expect significant differences of the relevant results for different flow durations, in the observed range.

In the reference case, since the confinement of the plasma is efficient, the thickness of the flow defines the thickness of the loop, in which the jet propagates (Fig.\ref{fig:inicond}a). The flow is defined solely by the velocity at the lower boundary of the domain. The density and temperature are the same as those of the medium from which it originates, i.e.  $n \sim 10^{11}$ cm$^{-3}$ and  $T \sim 10^{4}$ K, respectively. We do not account for possible loop expansion from the chromosphere to the corona \citep{2014arXiv1402.0338G}.

\begin{figure*}[!t]
	\centering
	\resizebox{\textwidth}{!}
	{
	\includegraphics[width=0.4\columnwidth]{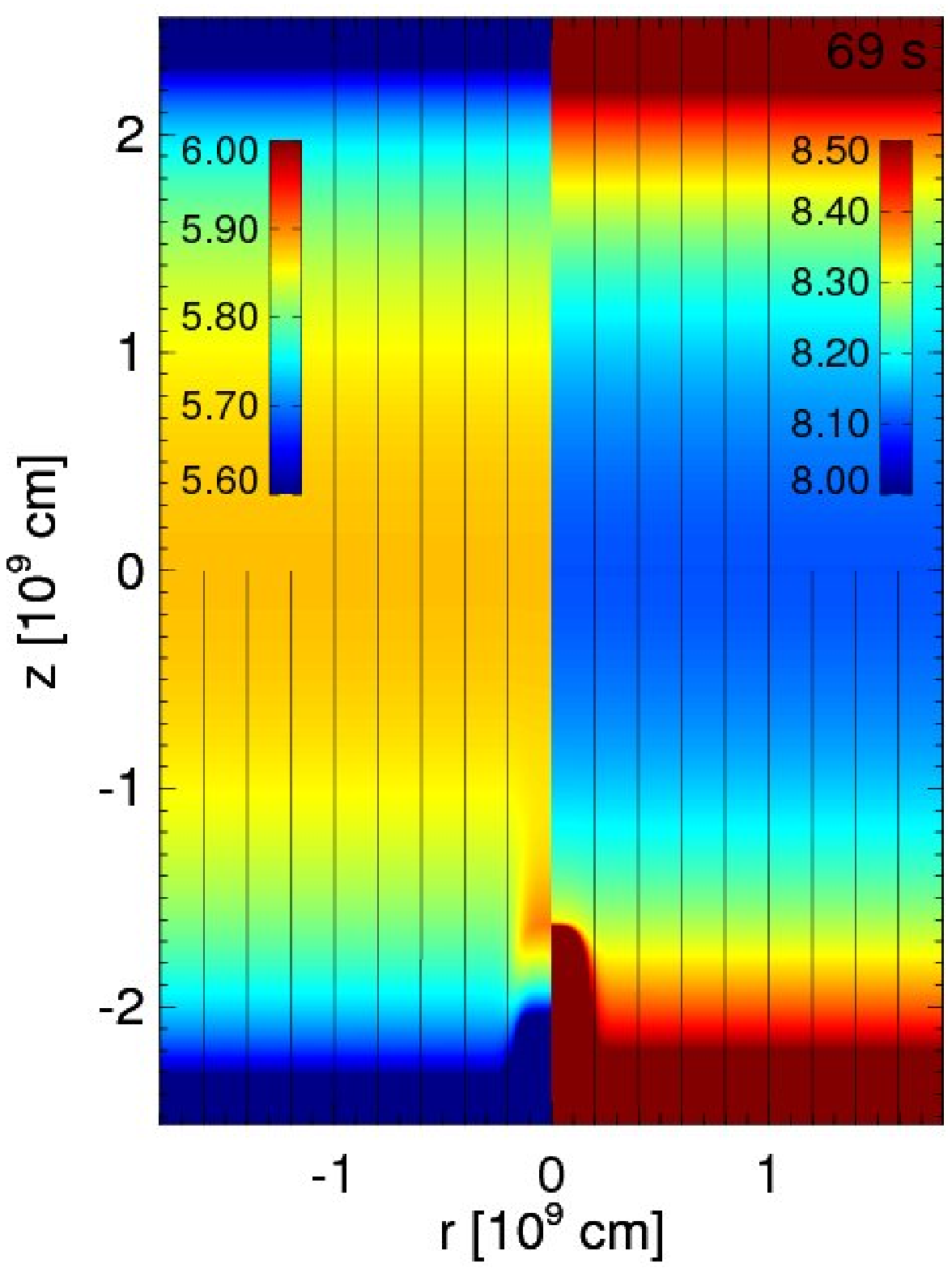}
	\includegraphics[width=0.4\columnwidth]{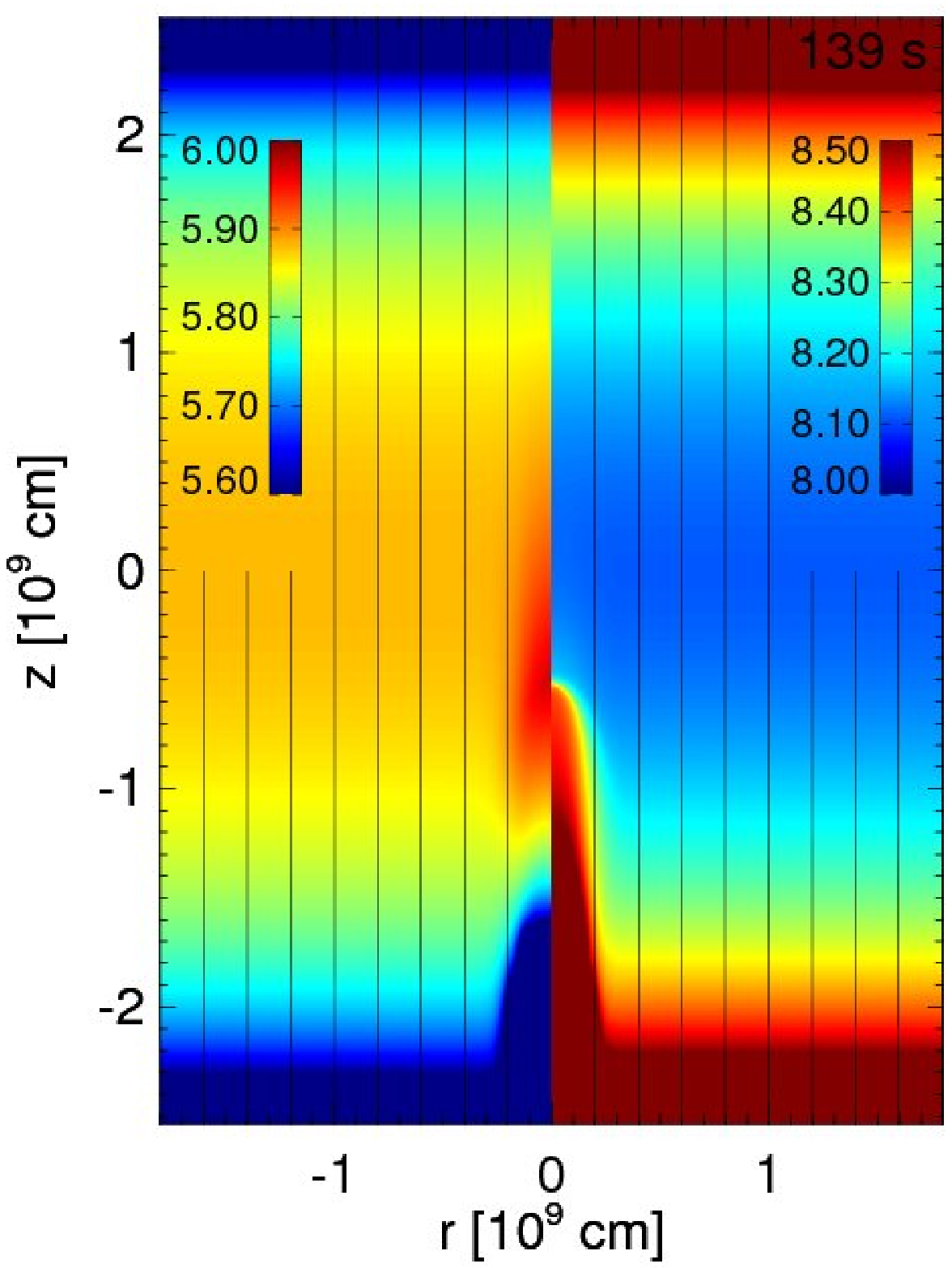}
	}
	\resizebox{\textwidth}{!}
	{
	\includegraphics[width=0.4\columnwidth]{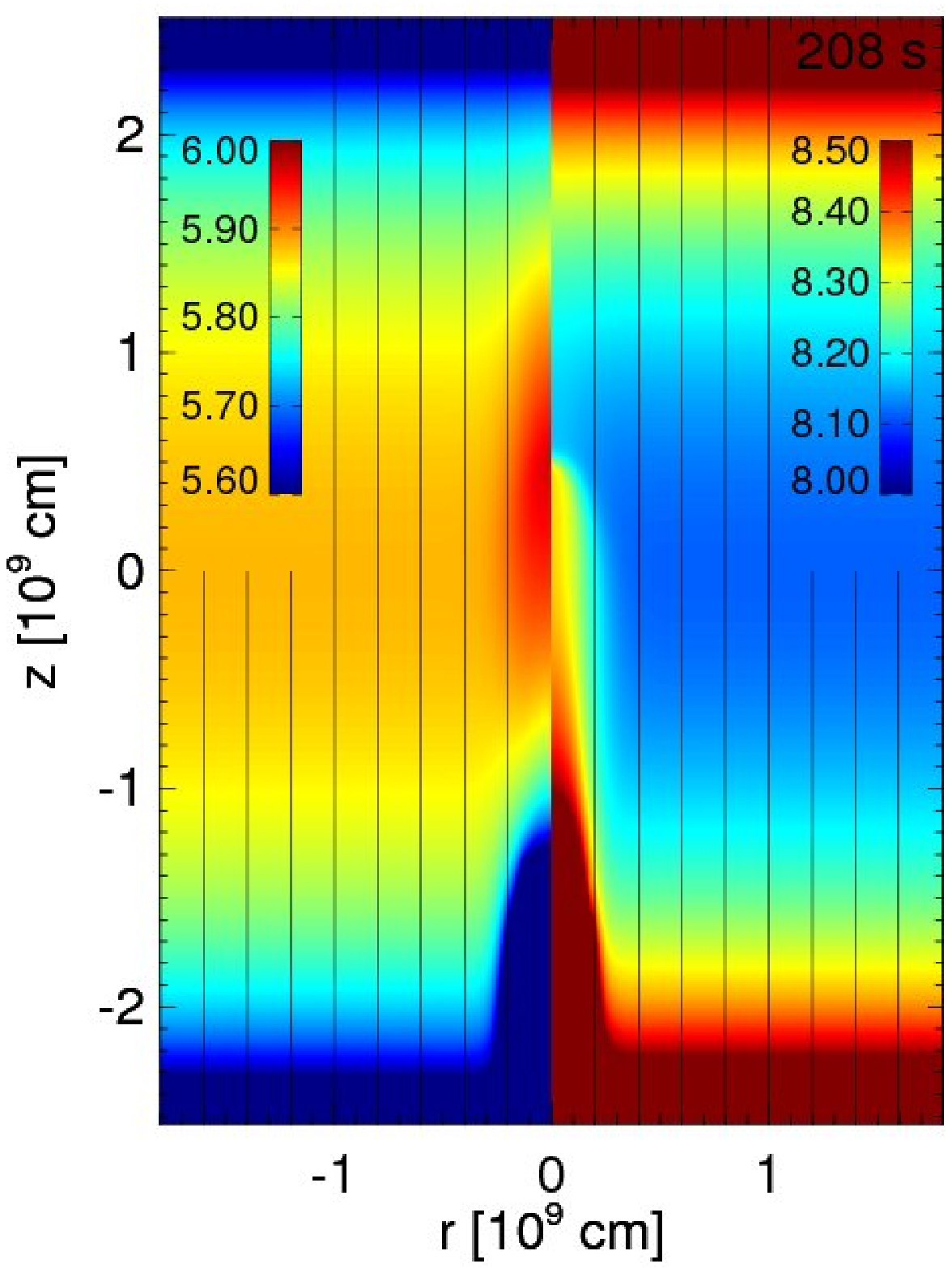}
	\includegraphics[width=0.4\columnwidth]{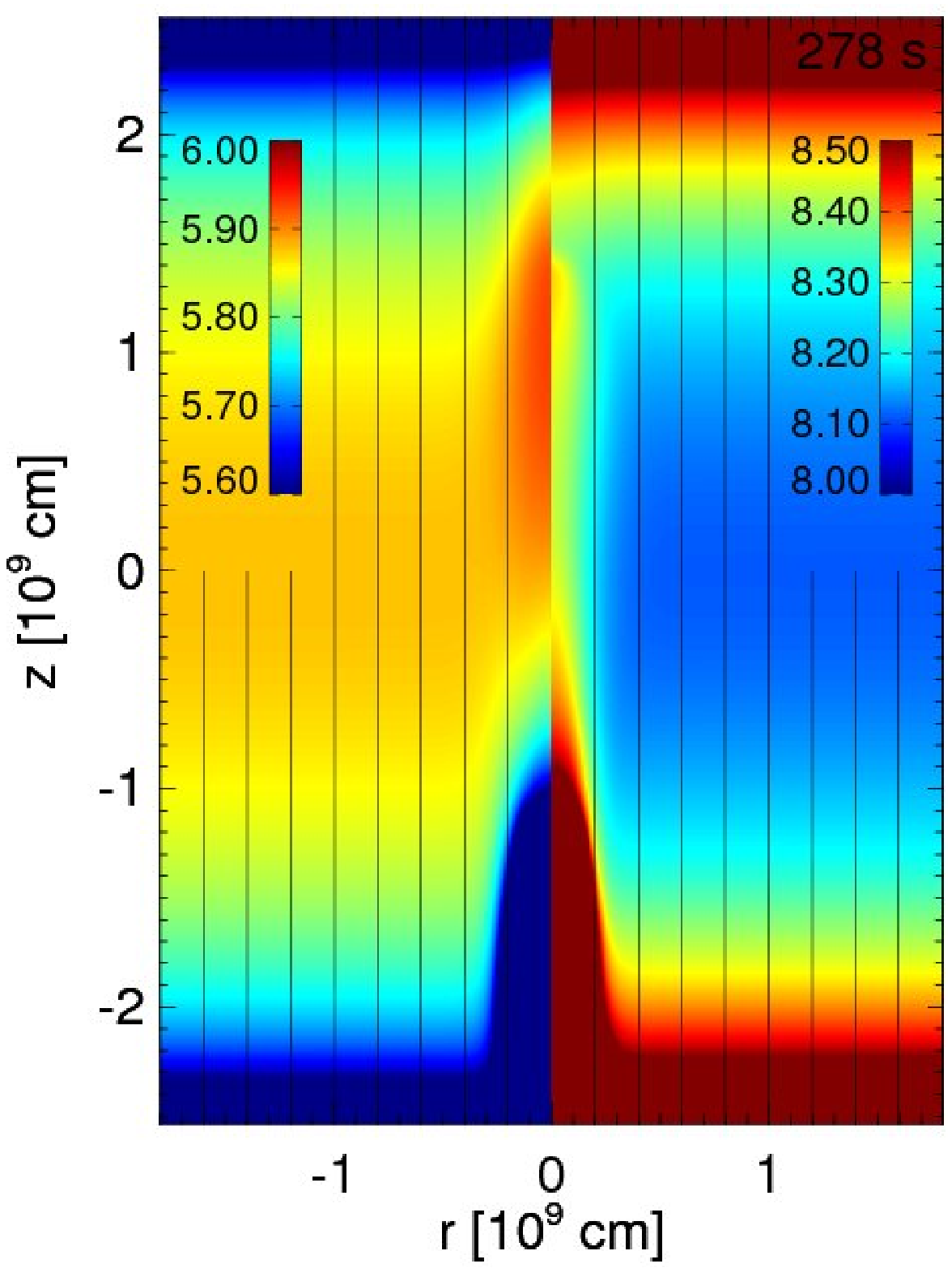}
	}
	\caption{Results of the reference 2-D MHD simulation: maps of temperature (left side) and density (right side) in a cross-section of the cylindrical domain at the labelled times (t = 69 s, 139 s, 208 s, 278 s, see on-line Supplementary movie for an animated version of this evolution).}
	\label{fig:mapsshock}
	\end{figure*}

\begin{figure*}[!t]
	\centering
	\includegraphics[scale= 0.7]{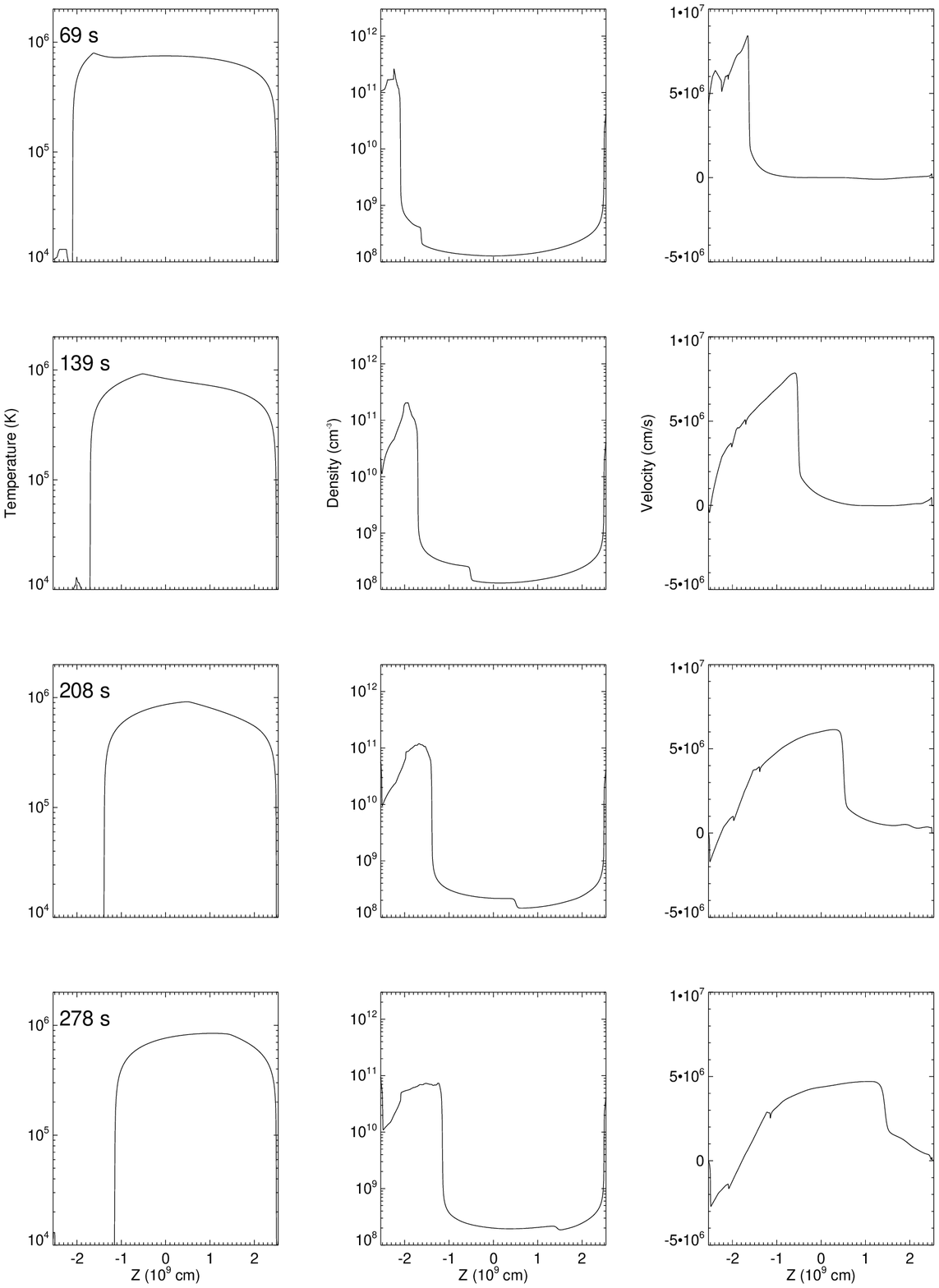}
	\caption{Results of the MHD simulation: temperature, density and velocity profiles along the central axis of the cylindrical domain at the labelled times.}
	\label{fig:plotB10v70n11}
	\end{figure*}

The evolution of the simulated jet is presented in Figs.\ref{fig:mapsshock} and \ref{fig:plotB10v70n11}. This work addresses the evolution of the coronal plasma outside of the jet, so we focus on the dynamics of the jet and we do not discuss the jet internal structure and its thermal evolution in detail. The speed value does not allow the jet to reach the apex of the loop, so after traveling a distance of $\sim2\times10^{9}$ cm (from the base of the loop) in 500 s, it then falls back to the chromosphere due to gravity. This distance corresponds to an altitude of $\sim1.6\times10^{9}$ cm with our curved loop gravity function, and is consistent with a simple ballistic estimate:

\begin{equation}
h=\frac{v^{2}}{2g_{\odot}}\sim  10^9 \qquad cm
\end{equation}

Fig.\ref{fig:mapsshock} shows maps of temperature and density through a cross-section of the cylindrical domain at four subsequent times (Fig.\ref{fig:inicond}a shows the last frame of Figs.\ref{fig:mapsshock} with a different, broader color scale).
	
At $t=69$ s, the cold jet has overcome the chromosphere, having travelled for a distance $\Delta z \sim 5 \times 10^{8}$ cm, and it is clearly visible as a blue bump in the temperature map.
%[**JK: It is very difficult to read the times in the plots. Perhaps they should be given in the caption as well.]
A shock front also propagates ahead of the cold temperature front and is visible as a red bump in the density map. A weaker thermal front, due to the efficient thermal conduction, is visible ahead of the shock, as a slight reddening along the central axis in the temperature map. At time $t = 139$ s, the jet is no longer being driven at the base. The fronts have all moved upwards, with different speeds that enlarge the distances between them. While the jet has advanced only to a height of  $\sim 10^{9}$ cm, the shock has almost reached the loop apex and the thermal front has a long tail that extends even farther. At $t = 208$ s, the jet has reached its maximum height and is now almost stationary due to the gravitational deceleration. The shock and the thermal front instead are still moving ahead to the other end of the loop. At $t = 278$ s, the thermal front has reached the other end of the loop, preceding the shock. As mentioned above, the magnetic field is not much affected by the propagation of the jet, that remains well collimated and confined, with a sharp flat front, as shown in Fig.\ref{fig:inicond}a. Some weak density and temperature halo is visible at the base of the corona in the late panels of Fig.\ref{fig:mapsshock}, emphasized by the saturation due to the expanded logarithmic scale. The post-shock plasma has some small expansion while linking to the cold jet on the back. The round shape of the jet front in Fig.\ref{fig:mapsshock} is only apparent and due to the saturated color scale (see Fig.\ref{fig:inicond}a).

Later the jet stops and eventually falls back downward to the surface, while the shock hits against the chromosphere at the other end of the loop. This later evolution (not shown) does not add interesting information to our discussion.

We focus now on some quantitative details. Figure \ref{fig:plotB10v70n11} shows profiles of temperature, density (logarithmic scale) and velocity along the central axis of the domain.

The density peak ($n \sim 2 \times 10^{11}$ cm$^{-3}$) marks the head of the cold jet, that moves rightwards in the plot. At $t = 69$ s its front is at $z \sim -2 \times 10^9$ cm and has moved to $z \sim -10^9$ cm at $t = 278$ s, with an average speed of $\sim 50$ km/s. Due to the large heat capacity of the jet, the hotter corona above is unable to heat the cold jet plasma by thermal conduction and the jet temperature remains mostly below $10^4$ K throughout its propagation. The density of the jet decreases gradually below $10^{11}$ cm$^{-3}$ as time progresses. The plasma slows down from $\sim 60$ km/s to $\sim 30$ km/s (it will stop completely later on and return back to the chromosphere).

At $t = 69$ s on the left side of the temperature profile ($z \approx -1.6 \times 10^9$ cm) we see a small cusp at about $10^6$ K. This cusp marks the shock front, that is more clearly visible as a jump in the density profile and as a sharp velocity peak (at 80 km/s). This front propagates rightwards along $z$ with an approximately constant speed $v_{sh} \sim 140$ km/s. At $t = 278$ s the shock front is nearly at the right end of the domain ($z \sim 1.5 \times 10^9$ cm). The density jump is somewhat reduced during the propagation, and also the plasma speed at the shock front decreases to about 60 km/s.

\begin{figure}[!t]
	\centering
	\includegraphics[scale= 0.7]{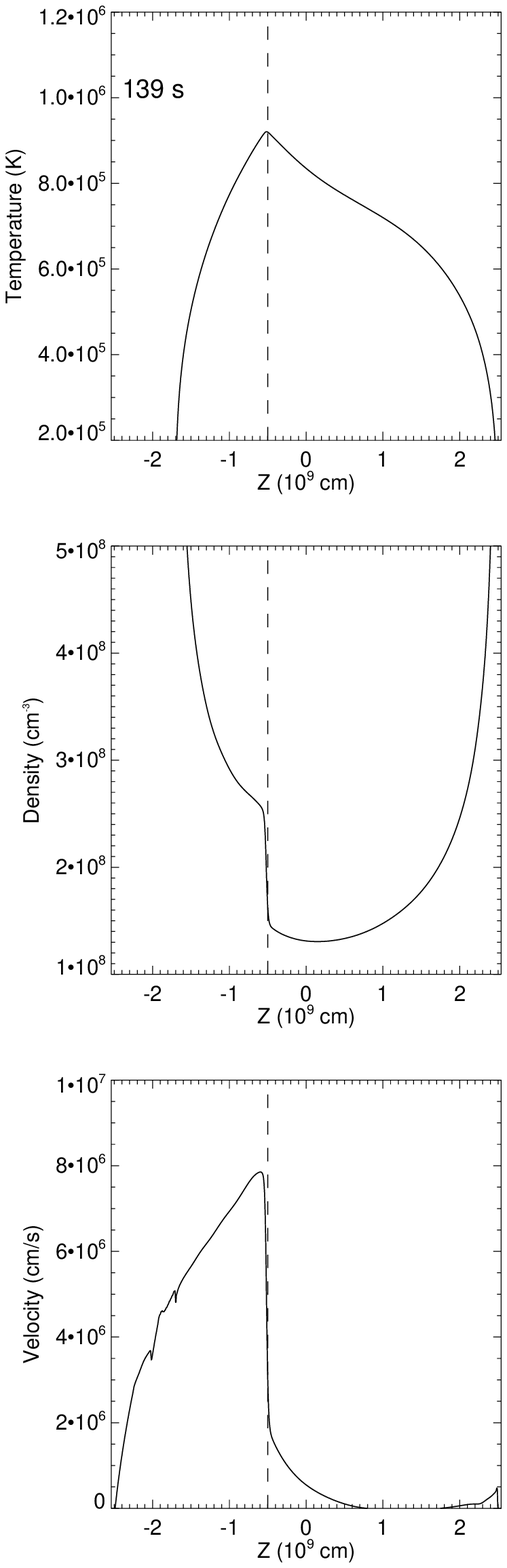}
	\caption{Detail of Fig.~\ref{fig:plotB10v70n11} at time t = 139 s (temperature, density and velocity). The position of the shock front is marked (vertical dashed line). }
	\label{fig:shock}
	\end{figure}

A detail of this evolution at time $t = 139$ s is shown in Fig.~\ref{fig:shock}. The shock front is visible very well in all three plots. On the sides of the temperature cusp we see that the temperature smoothly links to the original temperature profile. This smooth trend is the thermal front that we see in Fig.~\ref{fig:mapsshock}. The jump in density between pre- and post-shock medium is approximately a factor of two. This scenario is consistent with the following estimates.
	
Since the efficient thermal conduction in the corona makes the shock nearly isothermal, the isothermal sound speed is the reference speed:

\begin{equation}
c_{s} \approx \sqrt{\frac{2 k_B T_{sh}}{\mu m_H}} \sim 100 ~~ {\rm km/s}
\end{equation}
where $k_B$ is the Boltzmann constant, and $T_{sh} = 8 \times 10^5$ K.

From the Rankine-Hugoniot shock conditions (e.g. \citealt{landau_fluid_1987}) we know that:

   \begin{equation}
   \frac{\rho_{2}}{\rho_{1}} = \frac{(\gamma+1)}{\frac{2}{M^{2}}+(\gamma-1)}
   \label{eq:shock}
   \end{equation}
where $\rho_{2}$ and $\rho_{1}$ are the post-shock and pre-shock plasma mass densities, $\gamma$ is the ratio of the specific heats ($\gamma = 1$ for an isothermal plasma) and $M$ is the Mach number:

   \begin{equation}
   M = \frac{v_{1}}{c_{s}}
   \end{equation}

where $v_{1}$ is the pre-shock plasma speed in the reference frame of the shock front. We measure $v_1 \approx 140$ km/s, i.e. $M \approx 1.4$. For $\gamma = 1$, we obtain ${\rho_{2}}/{\rho_{1}} \approx 2$ that is fully compatible with the density jump in Fig.~\ref{fig:shock}.

Moreover, again from Rankine-Hugoniot shock conditions we have $v_2 = (\rho_1/\rho_2) v_1 \approx 0.5 v_1 \approx 70$ km/s in the shock frame and 140 - 70 = 70 km/s in the rest frame.
This is in good agreement with the peak velocity shown in Fig.~\ref{fig:shock}.
The propagation of the jet and of the shock alters the loop equilibrium. At t = 278 s, the shock has travelled all along the loop. At the same time, the length of the coronal part of the loop has effectively decreased because one low part of it is now occupied by the chromospheric jet. We estimate that the loop half length decreases from $L \sim 2.5 \times 10^9$ cm to $\sim 1.8 \times 10^9$ cm. According to the loop scaling laws \citep{1978ApJ...220..643R} we would expect that, at equilibrium and with no change of steady heating, a loop shortening corresponds to  (somewhat less than linear) reductions of the pressure and temperature. Instead, because of the shock, at t = 278 s, the plasma pressure has increased from $\sim 0.03$ dyne cm$^{-2}$ to $\sim 0.04$ dyne cm$^{-2}$, and the temperature from $\sim 0.7$ MK to $\sim 0.9$ MK. Therefore, if we interpret this result in terms of equilibrium conditions, we might say that a conversion of kinetic into thermal energy associated with the shock corresponds to an additional heating input into the loop.

For different jet speeds, the solutions are qualitatively very similar; only we obtain slower or faster shock propagation (on the order of few tens of seconds), and a smaller or higher density jump at the shock ($\pm 10-20$\%) for slower or faster jet speeds, respectively. For higher density of the jet, the shock moves at the same speed as for lower density, but the shock jump is about 20\% higher in density.

	\begin{figure}[!t]
	\centering
	\includegraphics[scale=0.7]{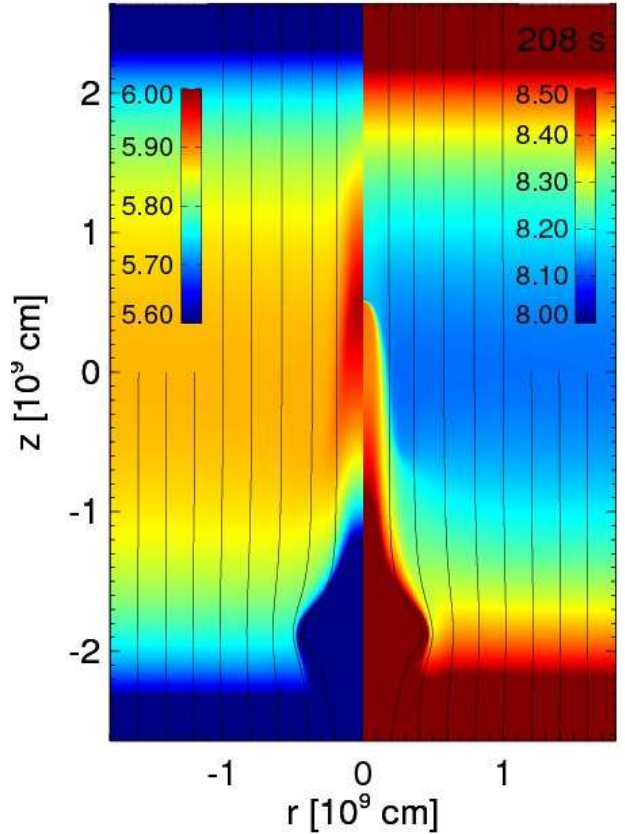}
	\caption{High $\beta$ MHD simulation: as in Fig.~\ref{fig:mapsshock} at time t =208 s (see on-line Supplementary movie for an animated version of this evolution).}
	\label{fig:highbeta}
	\end{figure}
In Fig.~\ref{fig:highbeta} we show a snapshot of the evolution that we obtain in the simulation with a higher value of plasma $\beta$. Although the confinement is no longer so efficient and the cold jet spreads out in the corona, the propagation of the shock ahead of the jet is little affected and remains very similar to that obtained for low $\beta$.

Finally, we comment also on the possible jet propagation in an initially hotter (1.5 MK) loop. A shock front still propagates along the loop, with a higher speed but quite a  low density jump ($\sim 30$\%).

\subsection{Synthetic emission}

Our interest here is mainly in the implications of this jet propagation for emissions in the EUV band. The jet itself is too cold to expect emission in this band. Instead the shocked plasma is at coronal temperature so we do expect that the density excess of the post-shock plasma leads to an emission excess in the EUV band with respect to the unperturbed atmosphere. In other words, a bright shock wave that propagates along the loop might be detectable in the EUV band.

To test this expectation we have synthesised the emission in the 171 \AA\ channel of the AIA instrument  \citep{2012SoPh..275...17L} on board the SDO mission. The filterband of this channel includes a very intense Fe IX line with a temperature of maximum formation of $\sim 10^6$ K. The emission in the selected channel is calculated as:

	\begin{equation}
	I_{171}(r,z,t) = G_{171}[T(r,z,t)] EM(r,z,t)
	\label{eq:EM_isotermo}
	\end{equation}
	where \[
    EM(r,z,t) = n_{e}^{2}(r,z,t) A_{pix}~~~,
	\]
$EM$ is the emission measure, $G_{171}$ is the sensitivity of the channel as a function of the temperature of the emitting plasma (available from the SolarSoftware package), and the AIA pixel area $A_{pix}$ is assumed. The depth along the line of sight is left as a free parameter. We have computed $I_{171}(r,z,t)$ for each grid cell of our 2D geometric domain, i.e. in the area across the central axis.

\begin{figure*}[!t]
	\centering
	\resizebox{\textwidth}{!}
	{
	\includegraphics[width=0.35\columnwidth]{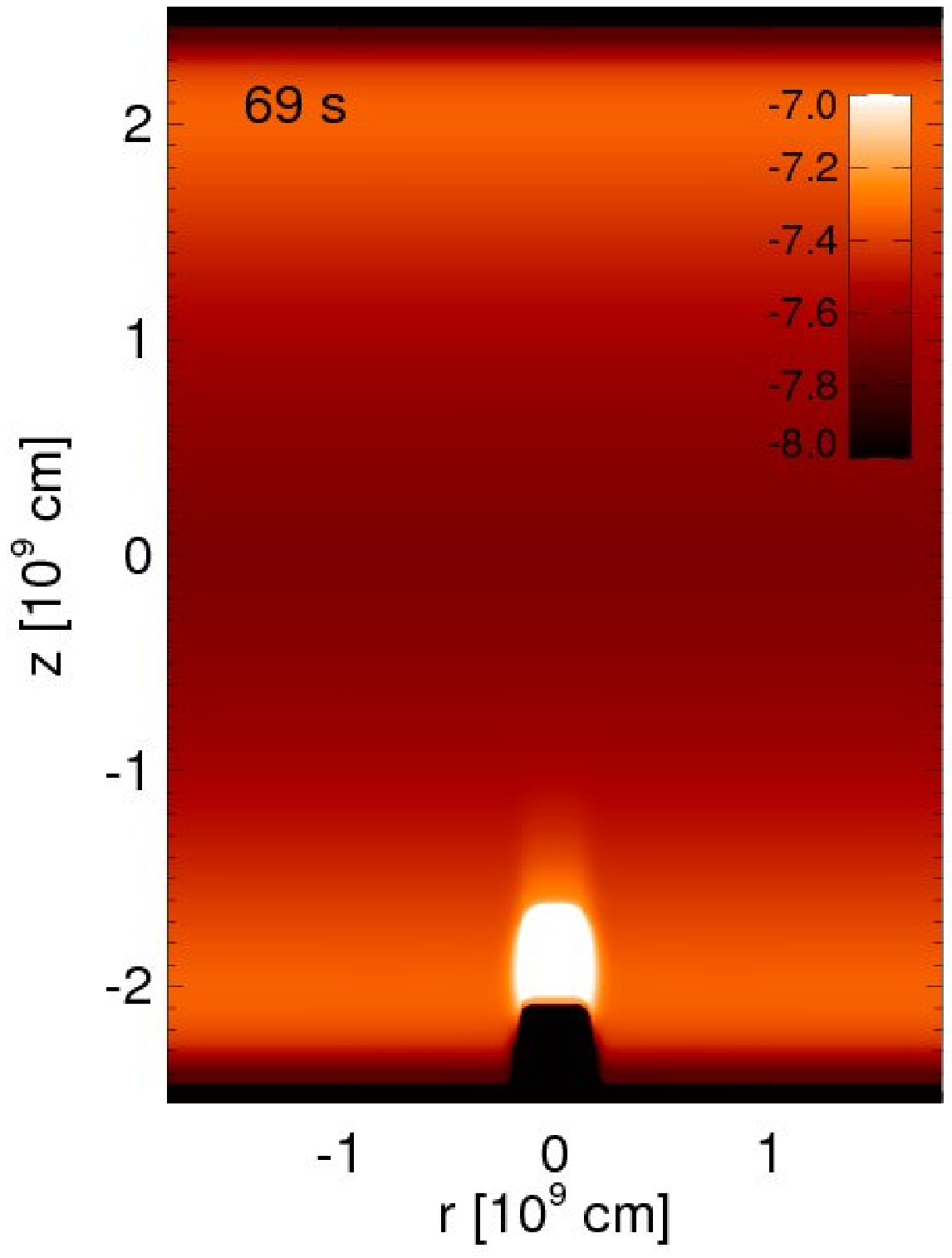}
	\includegraphics[width=0.35\columnwidth]{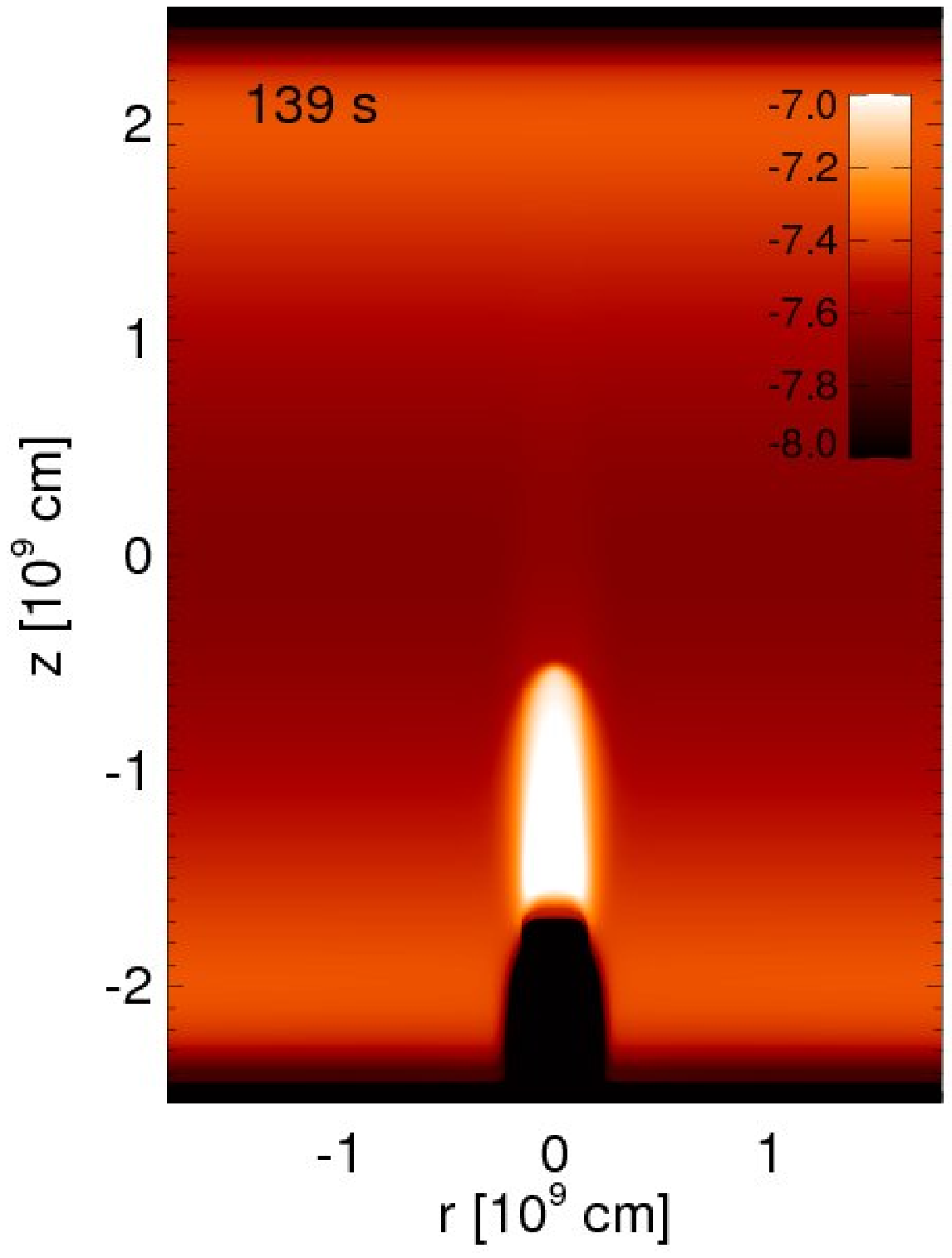}
	}
	\resizebox{\textwidth}{!}
	{
	\includegraphics[width=0.35\columnwidth]{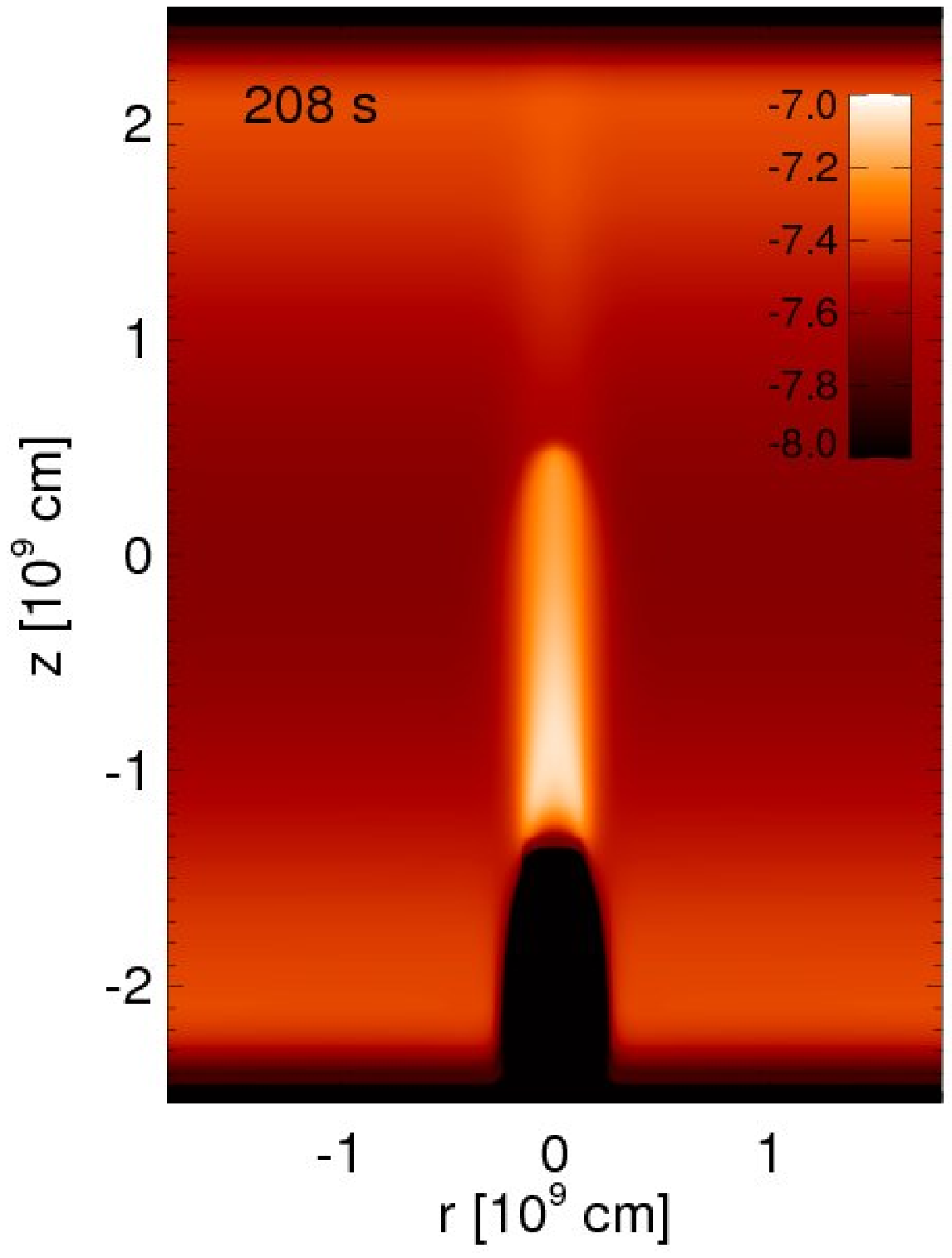}
	\includegraphics[width=0.35\columnwidth]{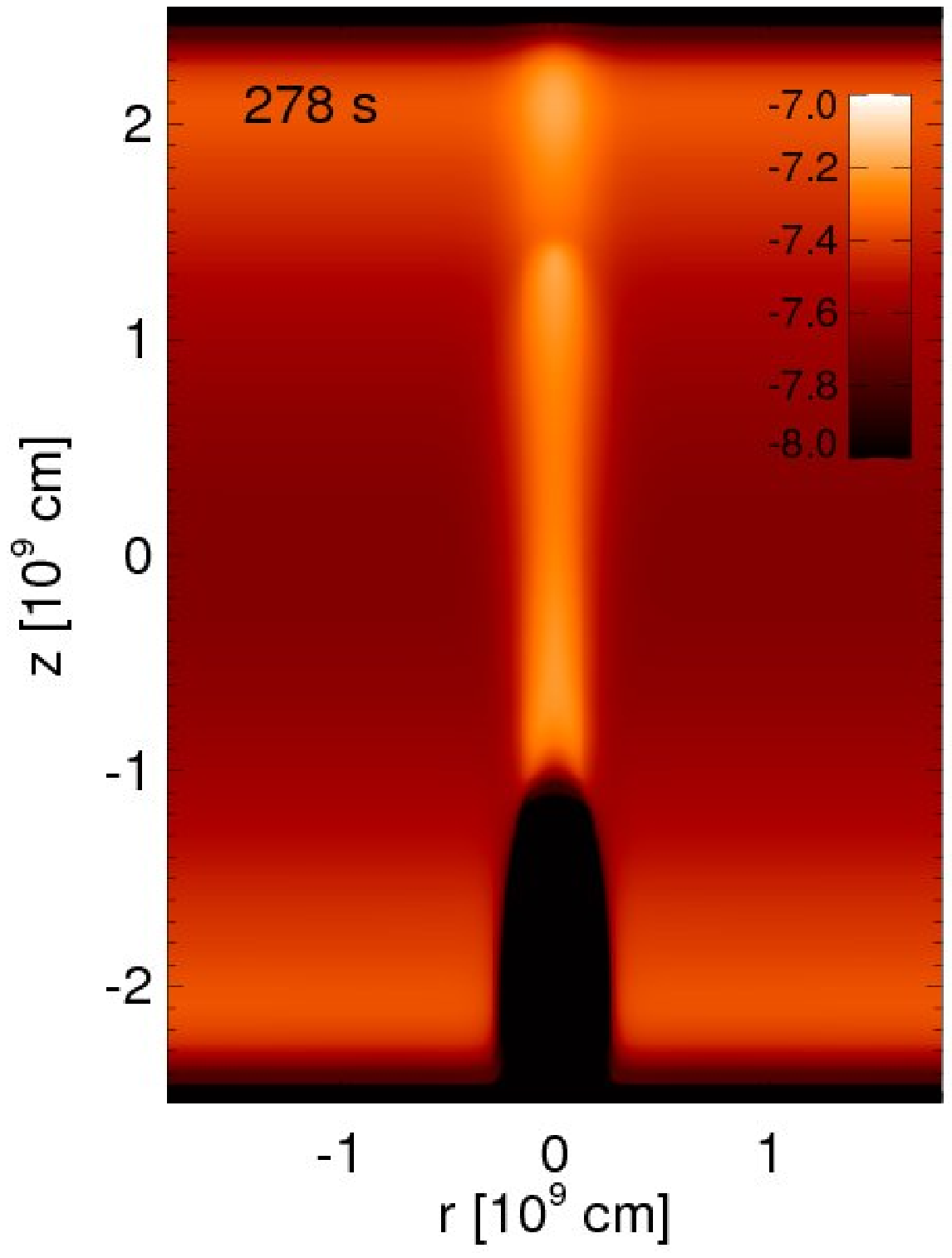}
	}
	\caption{Maps of synthesized emission (DN cm$^{-1}$ s$^{-1}$ pix$^{-1}$) in the 171 \AA\ channel of AIA instrument on board of SDO mission. We show a section of the computational domain at the same times of Fig.~\ref{fig:mapsshock}  (see on-line Supplementary movie for an animated version of this evolution).}
	\label{fig:mapsyntemis}
	\end{figure*}

\begin{figure}[!h]
	\centering
	\includegraphics[scale= 0.4]{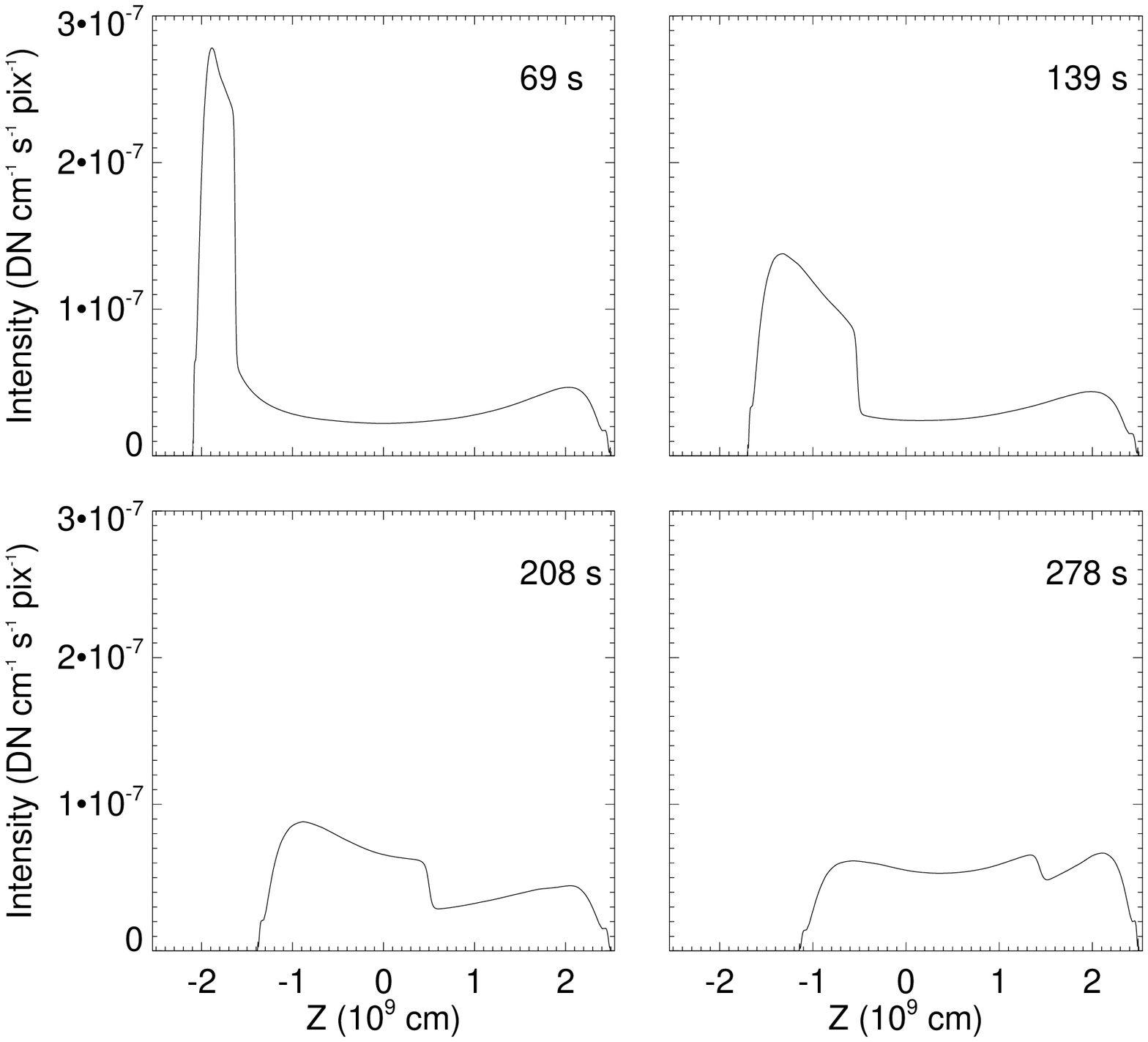}
	\caption{Emission profiles along the central axis of the cylindrical domain at the same times as Fig.~\ref{fig:mapsyntemis}.}
	\label{fig:plotemisZ}
	\end{figure}

Figs.~\ref{fig:mapsyntemis} and ~\ref{fig:plotemisZ} show four maps of the emission detectable in the 171 \AA\ channel at the same times and in a domain cross-section as Fig.~\ref{fig:mapsshock}, and the related profiles along the central $z$ axis, respectively, for the reference model.

In Fig. \ref{fig:mapsyntemis} we clearly see the brightening driven by the shock propagation. We can see the bright front propagating along the loop. This brightening is more intense at the beginning and then progressively fades as the shock moves along the coronal loop. The cold jet is the slower black bump.  At $t=278$ s we see a faint distinct halo on the opposite edge, which marks the presence of the weak thermal front. This effect would hardly be detected in the observations. In Fig.~\ref{fig:plotemisZ} we see that, initially, the emission of the post-shock medium is greater by about a factor 5 than the pre-shock medium (the upper part of the loop), but it is only a slight excess at time $t = 278$ s. This is about 3 minutes after the jet injection has ceased.

If we consider that the line-of-sight thickness of the bright moving front might be of the order of $10^7 - 10^8$ cm, we can estimate that the expected count rate is of the order of 1-10 DN s$^{-1}$ pix$^{-1}$, that may be measured in actual AIA observations. We remark that the apparent speed of the moving (shock) front is about 140 km/s, whereas the actual post-shock speed (measurable from Doppler shifts) would be less than 100 km/s.
In the other simulations, we do not see but some quantitative differences in the shock speed and front brightness, higher for higher jet speed and density. In the case of the hotter loop, the loop is already EUV bright since the beginning, and the shock perturbs the emission for a very short time and to a much lower extent, thus making the effect of the propagation much less observable.

\section{Discussion and conclusions}
\label{sec:discuss}

In this work we investigated chromospheric flows ejected upwards into a coronal loop. We considered a flow at a speed compatible with that measured for the so-called "type II spicules". The aim is to address their observed spatial and temporal correspondence with local brightenings in the extreme ultraviolet band, emitted by plasma typically at coronal temperatures. We considered a model of a simplified magnetic flux tube, considering a uniform magnetic field and a complete solar atmosphere from the chromosphere to the corona, and including all the physical terms of interest, in particular, gravity, radiative losses, thermal conduction along the field lines, the magnetic induction. The model solves numerically the magnetohydrodynamic equations in two-dimensional cylindrical geometry, implemented in the PLUTO parallel code. The chromospheric flow is set at $10^{4}$ K and ejected upwards at a speed of 30-90 km/s, causing the formation of a shock that precedes it. We consider a realistic jet duration of about 1 minute. The stream does not have sufficient initial impulse to reach the apex of the loop, while the shock front sweeps the entire loop, resulting in a compression of the coronal loop plasma where it propagates. The temperature of post-shock plasma increases to about $1 \times 10^6$ K and the density by a factor two. Both of these factors lead to a substantial increase of the plasma emission, just in the EUV band around 171\AA, where the correspondence with type II spicules is observed. This therefore suggests that at least part of the observed evidence can be produced by the formation of these shock fronts after chromospheric flows.

Our model assumptions should not affect the generality of this result. Our highly efficient thermal conduction, while typical of low corona conditions, can only lead to underestimate the predicted brightness of the post-shock plasma. Less efficient conduction would approach adiabatic conditions and therefore increase the density jump (up to a factor 4) at the shock front. We do not expect significant differences of the jet evolution for different jet widths. Also, we do not expect large differences in the presence of significant loop expansion in the transition region, because the shock would equally propagate in a corona with a constant cross-section. If any, the absolute intensity might increase because the thickness of the emitting region would increase along the line of sight.
We also tested that different jet speed and density only cause some quantitative differences, thus making the overall result robust in the observed ranges.

We emphasize that, while this mechanism might explain EUV observations of spicules, it does not explain the corona at large. Our model begins with a corona that is maintained at $T \sim 8 \times 10^{5}$ K by a steady coronal heating that is unrelated to the spicule. The primary effect of the shock is to compress the pre-existing hot plasma so that it becomes brighter. The shock raises the temperature only slightly to approximately $10^{6}$ K. Note that the density, emission measure, and therefore brightness of the compressed plasma depend on the initial density in the loop. If the initial plasma were less dense, due to a weaker coronal heating rate, then there would be less material accumulating ahead of the jet, and the speed of the shock front would be reduced (closer to that of the jet). The sound speed of the initial equilibrium would also be slower, but only very slightly, since it varies as $n^{1/4}$ (using the loop scaling laws in \citealt{1978ApJ...220..643R}). Consequently, the Mach number and therefore the shock compression ratio would be smaller, and the EUV emission would be fainter. On the other hand, modeling the jet propagation inside a loop already hot ($\sim 1.5$MK) since the beginning shows much less visible effects.

We note that the scenario studied here is different from that in \cite{2012JGRA..11712102K}. Our jet is not subjected to any extra heating, whereas Klimchuk looked at the case where the tip of the dense spicule is itself heated to coronal temperatures during the ejection.

In conclusion, the study presented here proposes the presence of shocks as candidates to explain the EUV emission in correspondence of the type II spicules. On a broader perspective, it is certainly a work that is leading the way to a series of other investigations of dynamic flows observed on the Sun in great detail by Solar Dynamics Observatory mission and others such as the Hinode mission. The MHD model developed for the present work can be applied to a wide variety of flows and their interaction with magnetic fields. These studies may lead to new knowledge that can even go beyond Solar Physics, as recently demonstrated with the connection between solar free falling flows and the accretion flows of star formation \citep{2013Sci...341..251R}.

% ***********************************************************

\acknowledgements{The authors thank the anonymous referee for suggestions. AP, FR and SO acknowledge support from Italian \emph{ Ministero dell'Universit\`a e Ricerca}. The work of JAK was supported by the NASA Supporting Research and Technology Program. PLUTO is developed at the Turin Astronomical Observatory in collaboration with the Department of Physics of the Turin University. We acknowledge the HPC facility (SCAN) of the INAF - Osservatorio Astronomico di Palermo, for the availability of high performance computing resources and support.
\emph{CHIANTI} is a collaborative project involving the \emph{NRL (USA)}, the \emph{Universities of Florence (Italy)} and
\emph{Cambridge (UK)}, and \emph{George Mason University (USA)}.}
%
% ***********************************************************
\bibliographystyle{aa}
\bibliography{corr}
\end{document}